\documentclass[
prc,%
10pt,%
final,%
notitlepage,%
oneside,%
twocolumn,%
nobibnotes,%
nofootinbib,
superscriptaddress,%
floatfix,%
showkeys,%
showpacs]%
{revtex4}

\usepackage{color} 
\usepackage{amsfonts} 
\usepackage{amsbsy} 
\usepackage{mathrsfs}
\usepackage{graphicx}
\usepackage{comment}

\def\lsim{\mathrel{\rlap{ \lower4pt\hbox{\hskip-3pt$\sim$}}
    \raise1pt\hbox{$<$}}} 
\def\gsim{\mathrel{\rlap{ \lower4pt\hbox{\hskip-3pt$\sim$}}
    \raise1pt\hbox{$>$}}} 
\def\scr#1{\mbox{\scriptsize #1}}

\begin{document}

\title{Light-nuclei production in heavy-ion collisions within a thermodynamical approach} 


%
\author{M. Kozhevnikova}\thanks{e-mail: kozhevnikova@jinr.ru}
\affiliation{Veksler and Baldin Laboratory of High Energy Physics,
  JINR Dubna, 141980 Dubna, Russia}
\author{Yu. B. Ivanov}\thanks{e-mail: yivanov@theor.jinr.ru}
\affiliation{Bogoliubov Laboratory of Theoretical Physics, JINR Dubna,
  141980 Dubna, Russia} 
  \affiliation{National Research Nuclear
  University "MEPhI",
  115409 Moscow, Russia} 
  \affiliation{National Research Centre
  "Kurchatov Institute", 123182 Moscow, Russia}

\begin{abstract}
We present results of simulations of light-nuclei production 
in relativistic heavy-ion collisions within updated
Three-fluid Hydrodynamics-based Event Simulator Extended 
by UrQMD (Ultra-relativistic Quantum Molecular Dynamics) final State interactions (THESEUS). 
The simulations were performed for Pb+Pb and Au+Au collisions in the collision energy range 
of $\sqrt{s_{NN}}=$  6.4--19.6 GeV. 
The light-nuclei production 
is treated within the thermodynamical approach on equal basis with hadrons.
The only additional parameter related to the light nuclei is the energy density of 
late freeze-out that imitates afterburner stage of the collision because the 
light nuclei do not participate in the UrQMD evolution. 
This parameter is fixed from the condition  
of the best reproduction of the proton transverse-momentum spectrum after the UrQMD afterburner 
by that at the late freeze-out. 
The updated THESEUS results in not perfect, but a reasonable reproduction of 
data on bulk observables of the light nuclei, 
especially their functional dependence on the collision energy  
and light-nucleus mass. 
Various ratios, $d/p$, $t/p$, $t/d$, and $N(t)\times N(p)/N^2(d)$, are also considered.
The directed flow of light nuclei turns out to be more involved. 
Apparently, it requires explicit treatment of the afterburner evolution of 
light nuclei that violates the kinetic equilibrium. 
Imperfect reproduction of the light-nuclei data leaves room for medium effects in produced light nuclei. 
\end{abstract}
\maketitle

\section{Introduction}

Interest in the light-nuclei production in heavy-ion collisions has been revived in connection with 
search for conjectured critical point in the QCD phase diagram. 
STAR experiment has already found possible indications of existence of the critical point
\cite{STAR:2020tga}. This observation was based on predicted \cite{Stephanov:2008qz}
peculiar dependence of scaled kurtosis of net-proton distribution
as a function of the collision energy. Besides, 
an enhanced production of light nuclei close to the critical point
with respect to a noncritical scenario is expected \cite{Shuryak:2019ikv,Shuryak:2020yrs,Sun:2020zxy}. 
This prediction is based on expectation that the
attractive part of nuclear potential becomes dominated by long-ranged critical mode of QCD. 
Abundant production of light nuclei may also result from 
formation of baryon clusters due
to spinodal decomposition associated with mechanically unstable region in the first-order 
phase transition 
\cite{Skokov:2008zp,Skokov:2009yu,Randrup:2009gp,Steinheimer:2012gc,Steinheimer:2019iso}. 
This spinodal clumping gets enhanced  
at the critical point, where fluctuations are too slow 
for the development of equilibrium mixed phase.

At present, there are several 3D dynamical models which include the coalescence mechanism of the light-nuclei production 
\cite{Russkikh:1993ct,Ivanov:2005yw,Liu:2019nii,Zhu:2015voa,Steinheimer:2012tb,Dong:2018cye,Sombun:2018yqh,Zhao:2020irc,Hillmann:2021zgj,Zhao:2021dka},
see also a recent review \cite{Oliinychenko:2020ply}. 
In the simplest version, the coalescence-based models 
deduce the relevant parameters from comparison 
with data on the light-nuclei production \cite{Russkikh:1993ct,Ivanov:2005yw}.
Therefore, their predictive power is restricted. 
Though the refined coalescence calculations are very successful in reproducing data 
in a wide range of collision energies \cite{Hillmann:2021zgj}. 
Advanced coalescence approaches involve the Wigner functions of light-nuclei
\cite{Liu:2019nii,Zhu:2015voa,Dong:2018cye,Sombun:2018yqh,Zhao:2020irc,Zhao:2021dka}
to calculate the coalescence parameters. 
The recently developed transport models, such as  
SMASH (Simulating Many Accelerated Strongly-interacting Hadrons)     
\cite{Weil:2016zrk,Oliinychenko:2018ugs,Staudenmaier:2021lrg},  
PHQMD (Parton-Hadron-Quantum-Molecular-Dynamics) 
\cite{Aichelin:2019tnk,Glassel:2021rod,Bratkovskaya:2022vqi} and a stochastic kinetic 
approach \cite{Sun:2021dlz}, 
treat light nuclei microscopically 
(so far, only deuterons in the SMASH \cite{Oliinychenko:2018ugs,Staudenmaier:2021lrg}) 
on an equal basis with other hadrons. 
However, these transport models also require a extensive additional input for treatment  
the light-nuclei production, albeit in a wide range of collision energies.

The thermodynamical approach does not need any additional parameters for treatment of the 
light-nuclei production. It describes the light nuclei in terms of temperatures and chemical potentials,
i.e. on the equal basis with hadrons. Therefore, its predictive power is the same for light nuclei and hadrons. 
This approach was realized within the statistical model \cite{Andronic:2005yp,Cleymans:2005xv}: 
deuteron midrapidity yields at the energies (from 7.7 to GeV 200 GeV) of the STAR Beam Energy
Scan (BES) at  the Relativistic Heavy-Ion Collider (RHIC) \cite{STAR:2019sjh,STAR:2022hbp}
are described fairly well by this model \cite{Andronic:2010qu,Vovchenko:2020dmv}, while the yield of tritium is
overestimated by roughly a factor of two \cite{Vovchenko:2020dmv,Zhang:2020ewj}.
The statistical model gives a similarly good description of not only the light nuclei 
but even hypernuclei and antinuclei 
at energies of the CERN Large Hadron Collider (LHC) \cite{Andronic:2017pug}. 
The apparent success of the thermal model is puzzling. It is hard to imagine
that nuclei exist in the hot and dense fireball environment. The temperature is much
higher than the binding energy and the system is quite dense, so that the inter-particle spacing
is smaller than the typical inter-nucleon distance in a nucleus. This puzzle is discussed 
in Refs. \cite{Shuryak:2020yrs,Oliinychenko:2018ugs,Mrowczynski:2020ugu}.

In view of the success of the thermal model, we have implemented the thermodynamic approach of the 
light-nuclei production into the updated THESEUS event generator \cite{Kozhevnikova:2020bdb}.
In this paper, we address the question of how well this thermodynamic approach can describe the data 
on the light-nuclei production, provided the bulk observables 
\cite{Ivanov:2012bh,Ivanov:2013wha,Ivanov:2013yla,Ivanov:2018vpw}
for protons are reasonably well reproduced by the  model  of the three-fluid dinamics (3FD).  
Note that the model involves no extra parameters (except for the late freeze-out energy density, 
see sect. \ref{Results}) related to the light nuclei. 
For this purpose we analyze the available data from NA49 \cite{Anticic:2016ckv} 
and STAR \cite{STAR:2022hbp,Zhang:2020ewj} collaborations.

\section{updated THESEUS} 
  \label{updated THESEUS}

The THESEUS event generator
was first presented and its applications to 
heavy-ion collisions were demonstrated in Refs.~\cite{Batyuk:2016qmb,Batyuk:2017sku}. 
The THESEUS is based on the 3FD model~\cite{Ivanov:2005yw,Ivanov:2013wha}
complemented by the UrQMD~\cite{Bass:1993em,Bass:1998ca} for the afterburner stage.
The output of the 3FD model, i.e. the freeze-out hypersurface, is recorded in terms of
local flow velocities and thermodynamic quantities. 
The THESEUS generator transforms the 3FD output
into a set of observed particles, i.e. performs the particlization. 

The 3FD  is designed to simulate heavy-ion collisions at energies 
of the BES-RHIC 
at the Brookhaven National Laboratory~(BNL),
CERN Super-Proton-Synchrotron~(SPS), 
the Facility for Antiproton and Ion Research~(FAIR) in Darmstadt and the
Nuclotron-based Ion Collider fAcility~(NICA) in Dubna.
 It takes into account counterstreaming of the leading baryon-rich matter at the early stage of
nuclear collisions. This nonequilibrium stage is modeled 
by the means of two counterstreaming baryon-rich fluids.  
Newly produced particles, which dominantly populate the midrapidity region,
are assigned to a so-called fireball fluid.  These fluids are governed  
by conventional hydrodynamic equations coupled by friction
terms, which describe the energy--momentum exchange between the fluids.

At present, three different equations of state (EoS's) are used
in the 3FD simulations: 
a purely hadronic EoS \cite{gasEOS} (hadr. EoS) and two EoS's
with deconfinement \cite{Toneev06}, i.e. 
an EoS with a first-order phase transition (1PT EoS) and one with a
smooth crossover transition (crossover EoS).  
At energies $\sqrt{s_{NN}}>$ 5 GeV, 
the deconfinement scenarios reveal definite preference~\cite{Ivanov:2013yqa}. 
Therefore, in the present work we consider only the 1PT and crossover EoS's. 

The 3FD and the original version of the THESEUS \cite{Batyuk:2016qmb,Batyuk:2017sku} 
calculate spectra of the so-called primordial nucleons, 
i.e. both observable nucleons and those bound in the light nuclei. This is done for 
the subsequent application of the coalescence model  \cite{Ivanov:2005yw,Ivanov:2017nae} 
for the light-nuclei production.

The light nuclei were included in the updated version of the THESEUS 
\cite{Kozhevnikova:2020bdb}.  
The list of the light nuclei includes the stable nuclei and 
and low-lying resonances of the $^4$He system, the decays of which contribute to the yields 
of stable species  \cite{Shuryak:2019ikv}, see Tab. \ref{tab:clusters}. 
The corresponding anti-nuclei were also included. 
\begin{table}[ht]
\begin{center}
\begin{tabular}{|c|c|c|}
\hline
Nucleus($E$[MeV]) & $J$ &  decay modes, in \% \\
\hline
\hline
$d$           & $1$ & Stable \\
$t$           & $1/2$ & Stable \\
$^3$He        & $1/2$ & Stable \\
$^4$He        & $0$ & Stable \\
$^4$He(20.21) & $0$ & $p$ = 100\\
$^4$He(21.01) & $0$ & $n$ = 24,  $p$ = 76\\
$^4$He(21.84) & $2$ & $n$ = 37,  $p$ = 63  \\
$^4$He(23.33) & $2$ & $n$ = 47,  $p$ = 53  \\
$^4$He(23.64) & $1$ & $n$ = 45,  $p$ = 55 \\
$^4$He(24.25) & $1$ & $n$ = 47,  $p$ = 50,  $d$ = 3 \\
$^4$He(25.28) & $0$ & $n$ = 48,  $p$ = 52\\
$^4$He(25.95) & $1$ & $n$ = 48,  $p$ = 52 \\
$^4$He(27.42) & $2$ & $n$ = 3,   $p$ = 3,   $d$ = 94 \\
$^4$He(28.31) & $1$ & $n$ = 47,  $p$ = 48,  $d$ = 5 \\
$^4$He(28.37) & $1$ & $n$ = 2,   $p$ = 2,   $d$ = 96 \\
$^4$He(28.39) & $2$ & $n$ = 0.2, $p$ = 0.2, $d$ = 99.6 \\
$^4$He(28.64) & $0$ & $d$ = 100 \\
$^4$He(28.67) & $2$ & $d$ = 100 \\
$^4$He(29.89) & $2$ & $n$ = 0.4, $p$ = 0.4, $d$ = 99.2 \\
\hline
\end{tabular}
\caption{Stable light nuclei and low-lying resonances of the~$^4$He system (from BNL properties of nuclides 
\cite{www.nndc.bnl}).
$J$ denotes the total angular momentum. 
The last column represents branching ratios of the decay channels, in per cents. The 
$p,n,d$ correspond to the emission of protons, neutrons, or deuterons, respectively.
}
\label{tab:clusters}
\end{center}
\end{table}
These nuclei are sampled similarly to other hadrons, i.e.\ accordingly to their 
phase-space distribution functions.
Nevertheless, there is an important difference in the treatment of light nuclei and other hadrons. 
While the hadrons
pass through the UrQMD afterburner stage after the the particlization, the light nuclei do not, just 
because the UrQMD is not able to treat them. This is a definite shortcoming because the light nuclei
are destroyed and re-produced during this afterburner stage 
\cite{Oliinychenko:2018ugs,Staudenmaier:2021lrg,Glassel:2021rod}.

\section{Results} 
  \label{Results}

To partially overcome the aforementioned problem of the afterburner stage for the light nuclei,
we imitate the afterburner effect by later freeze-out for light nuclei. 
For this imitation we need to estimate the suitable late freeze-out. This we do   
by fitting the afterburner effect for protons by means of the late freeze-out. 
We choose protons because they are closely related to the light nuclei. 
The basic idea behind such imitation is as follows. If the afterburner effect for protons 
can be imitated by the late freeze-out, we anticipate that the same late freeze-out 
can do the same for light nuclei.

\begin{figure}[!tbh]
  \includegraphics[width=.37\textwidth]{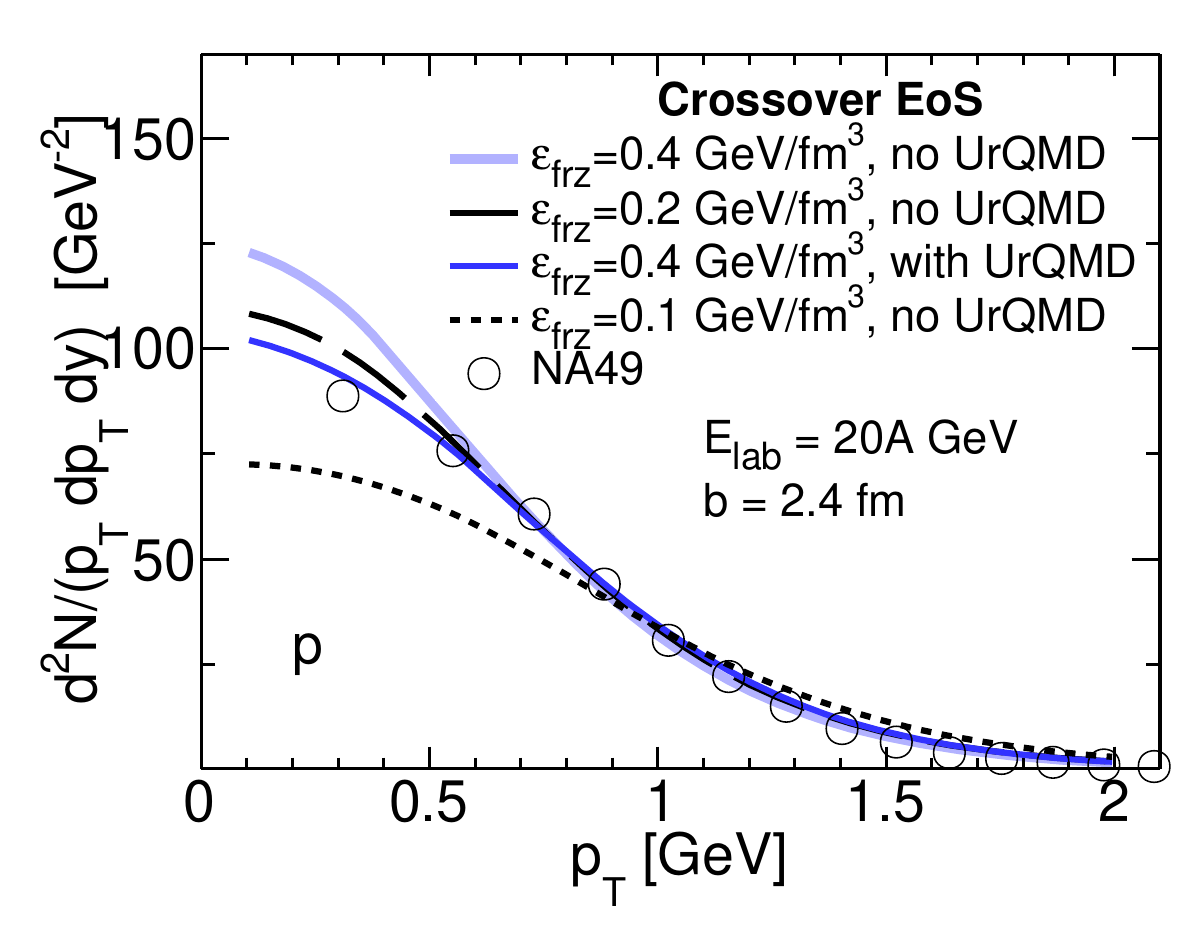}
  \includegraphics[width=.37\textwidth]{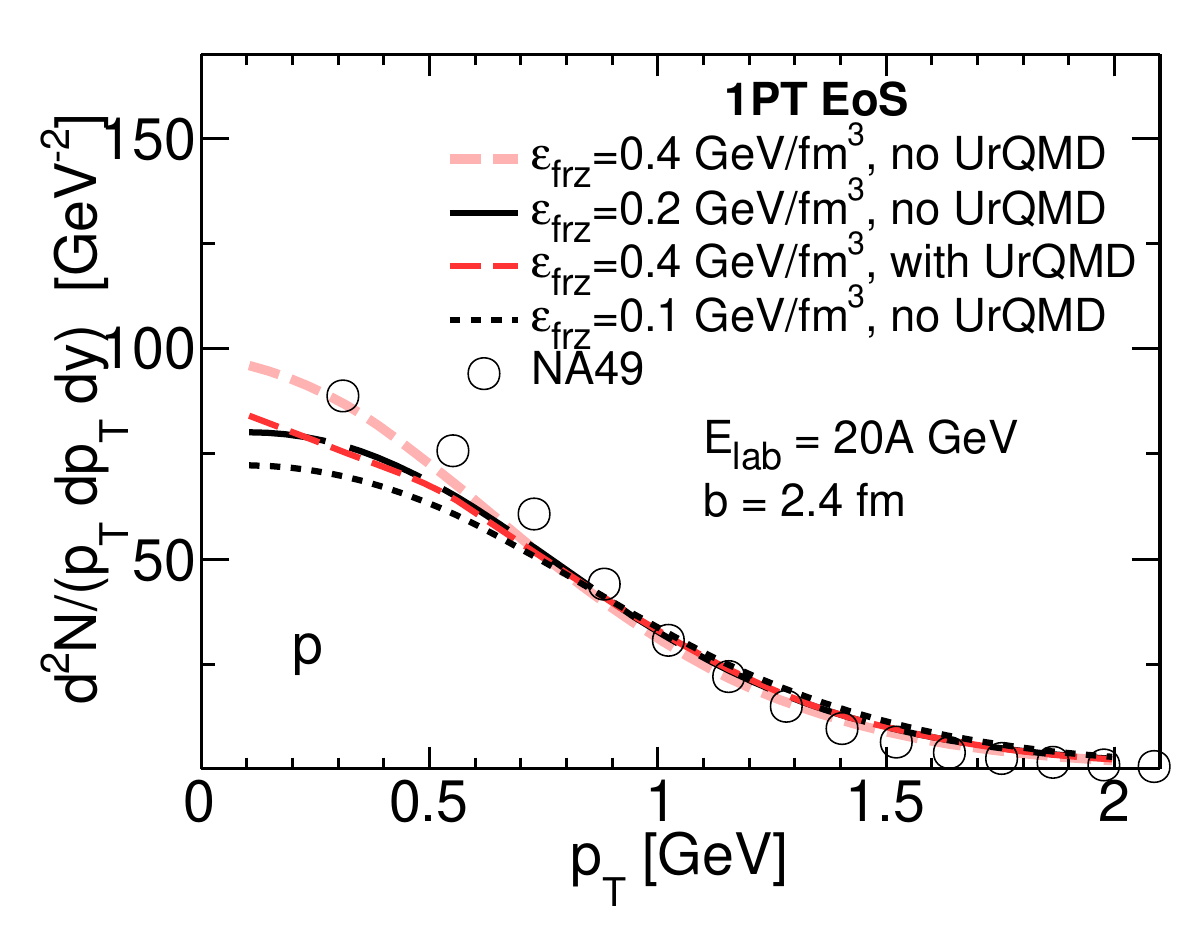}
  \caption{(Color online)
Transverse-momentum spectra of protons  
in central Au+Au collisions at collision energy of 
$E_{\rm lab}=$ 20$A$  GeV 
calculated with the crossover (upper panel) and 1PT EoS's (lower panel). 
Results of the THESEUS simulations (without the subsequent UrQMD afterburner)
based on the 3FD calculations with different 
freeze-out energy densities $\varepsilon_{\rm frz}=$ 0.1, 0.2 and 0.4 GeV/fm$^3$ are shown. 
The conventional for the 3FD results with $\varepsilon_{\rm frz}=$ 0.4 GeV/fm$^3$ 
and the subsequent UrQMD afterburner are also presented.   
Experimental data are from the NA49 collaboration  \cite{NA49:2004iqm}.
}		
\label{fig:pT_p_20AGeV}
\end{figure}
\begin{figure}[!tbh]
  \includegraphics[width=.37\textwidth]{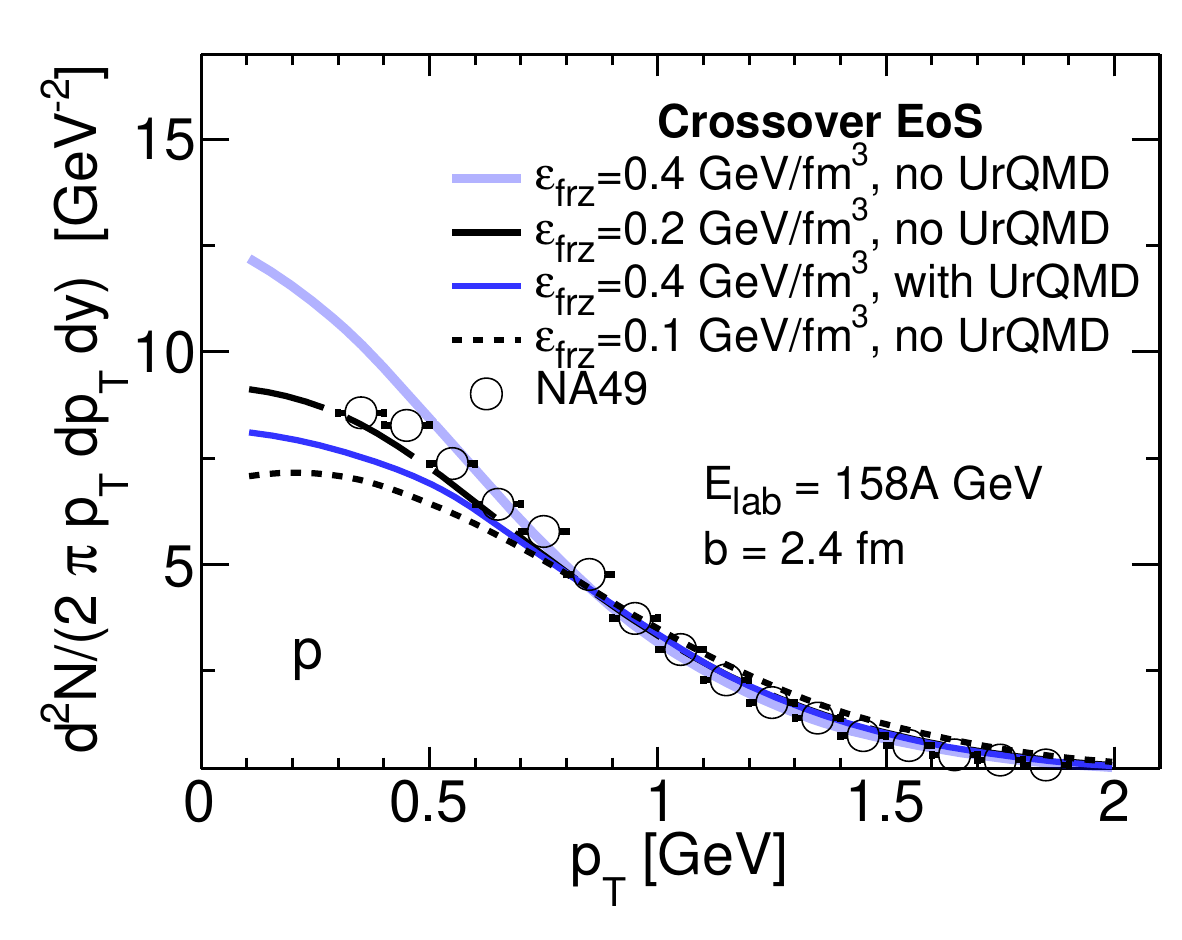}
  \includegraphics[width=.37\textwidth]{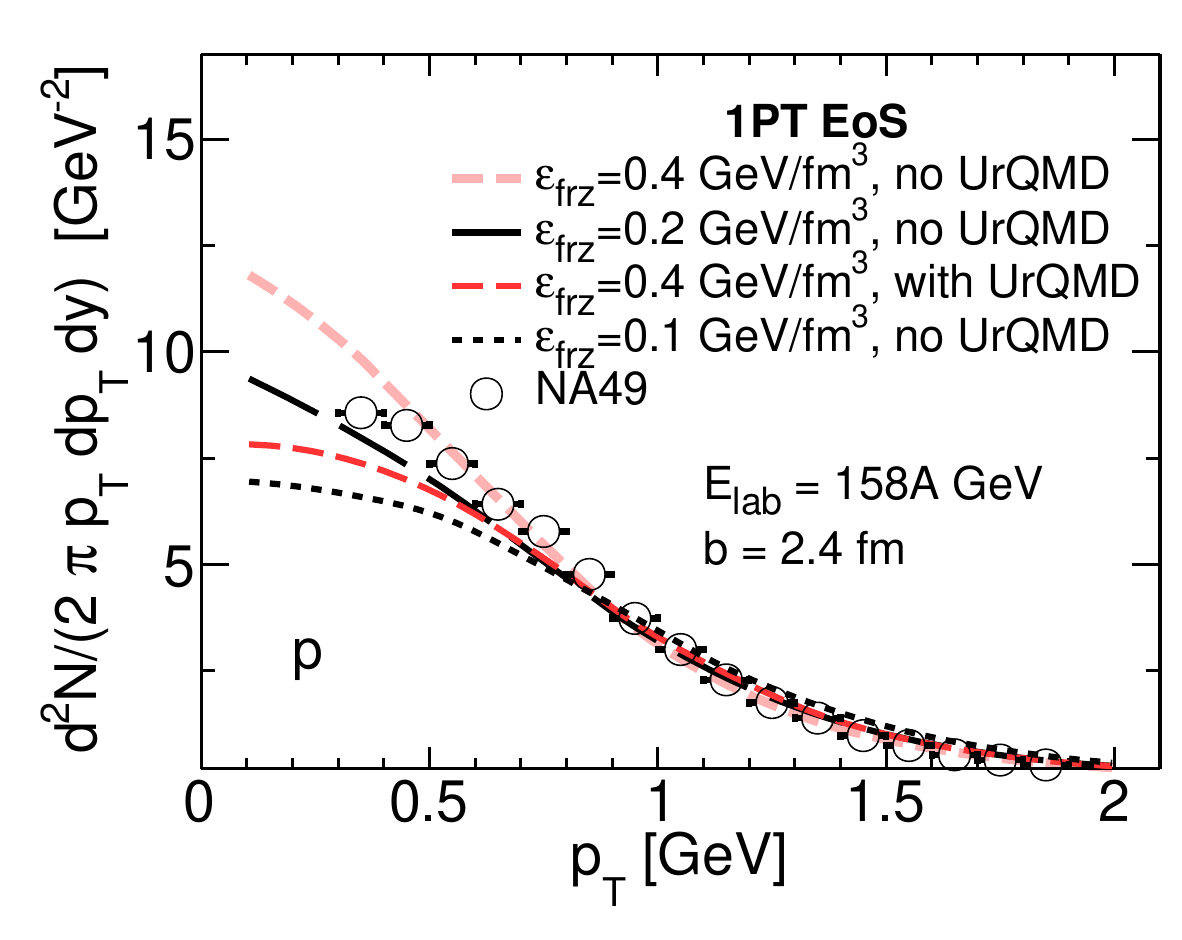}
  \caption{(Color online)
The same as in Fig. \ref{fig:pT_p_20AGeV} but for $E_{\rm lab}=$ 158$A$ GeV.  
Experimental data are from the NA49 collaboration  \cite{NA49:2007gga}.
}		
\label{fig:pT_p_158AGeV}
\end{figure}

In the 3FD calculations a differential, i.e.
cell-by-cell, freeze-out is implemented \cite{Russkikh:2006aa}. 
The freeze-out procedure starts when the local energy density drops down 
to the freeze-out value $\varepsilon_{\scr{frz}}$,   
which is conventionally taken to be 0.4 GeV/fm$^3$ for all collision energies 
and centralities. 
The freeze-out criterion is checked in the analyzed cell and in eight  
surrounding cells. If the criterion is met in all cells and
if the analyzed cell is adjacent to the vacuum 
(i.e if at least one of the surrounding cells is ``empty''\footnote{Frozen-out 
cells are removed from the hydrodynamical evolution.}), 
then the considered cell is counted as frozen out. The latter condition 
prevents formation of bubbles of frozen-out matter inside the
still hydrodynamically evolving matter. Thus, the actual energy density
of frozen-out cell turns out to be lower than $\varepsilon_{\scr{frz}}$. 
Therefore, $\varepsilon_{\scr{frz}}$ has a meaning of a ``trigger'' value
that indicates possibility of the freeze-out.
This freeze-out pattern is similar to the
process of expansion of a compressed and heated fluid into vacuum, 
mechanisms of which were studied
both experimentally and theoretically, see discussion in Ref. \cite{Russkikh:2006aa}. 
The freeze-out is associated with evaporation from the surface of the
expanding fluid.

In Figs. \ref{fig:pT_p_20AGeV} and \ref{fig:pT_p_158AGeV}, 
transverse-momentum spectra of protons  
in central Au+Au collisions at collision energies of 
$E_{\rm lab}=$ 20$A$ and 158$A$  GeV are shown. 
These spectra are calculated by means of  
the THESEUS simulations without the subsequent UrQMD afterburner, similarly to 
the light-nuclei simulations, 
based on the 3FD calculations with different 
freeze-out energy densities $\varepsilon_{\rm frz}=$ 0.1, 0.2 and 0.4 GeV/fm$^3$. 
The conventional for the 3FD results with $\varepsilon_{\rm frz}=$ 0.4 GeV/fm$^3$ 
and the subsequent UrQMD afterburner are also presented. 
The results are presented in linear scale in order to better resolve the low $p_T$
region, which is mostly affected by the afterburner effect \cite{Song:2010aq}.

As seen from Figs. \ref{fig:pT_p_20AGeV} and \ref{fig:pT_p_158AGeV},
the late freeze-out with the energy density $\varepsilon_{\rm frz}=$ 0.2 GeV/fm$^3$  
approximately reproduces 
the afterburner effect in midrapidity proton $p_T$ spectra at both collision energies and 
in different EoS scenarios. The reproduction of the high-$p_T$ spectra is also good, though 
it is hardly seen in the linear scale. 
We avoid fine tuning of the late freeze-out because this 
is only an imitation of the afterburner.

\begin{figure}[!tbh]
  \includegraphics[width=.33\textwidth]{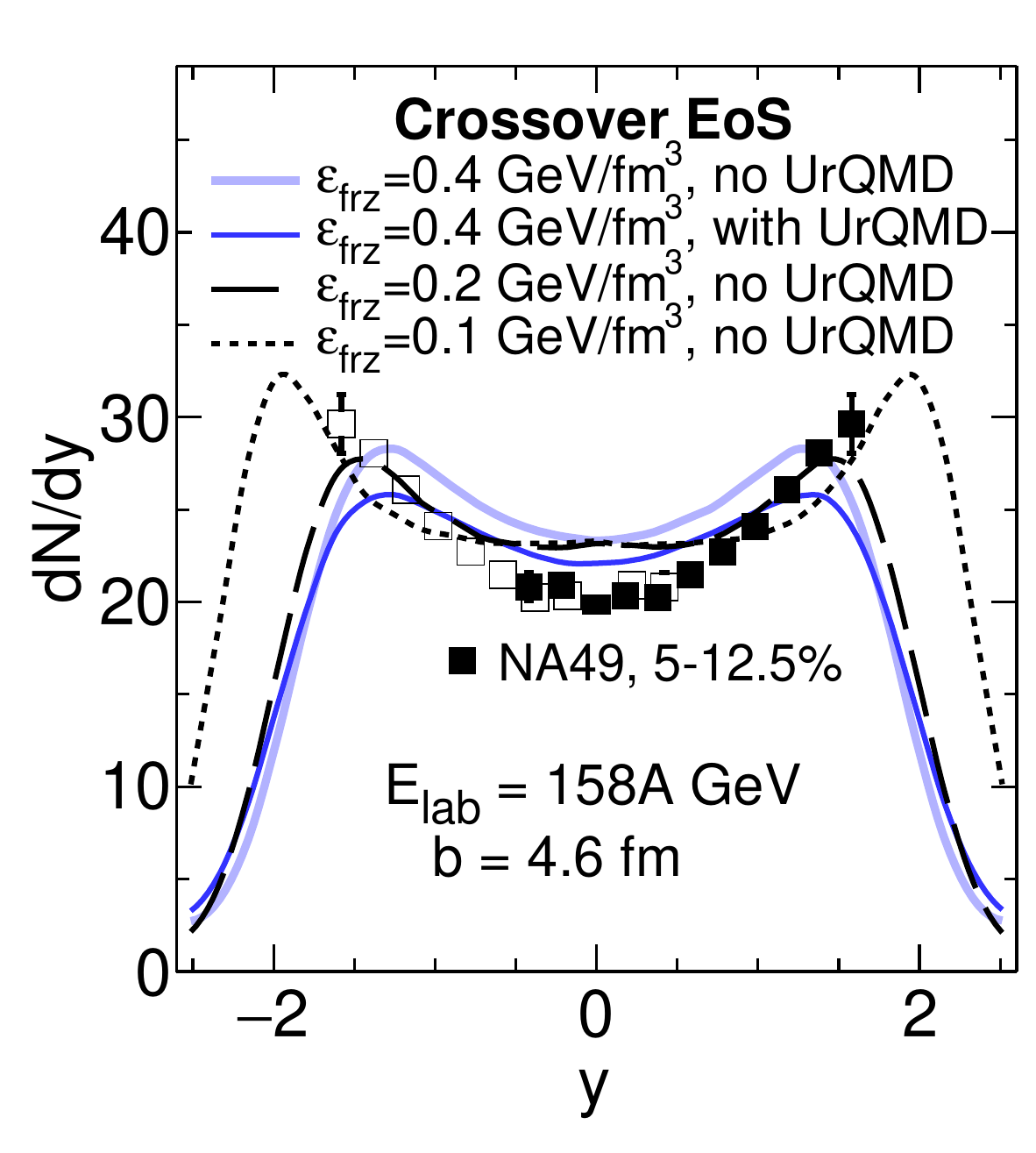}
  \includegraphics[width=.33\textwidth]{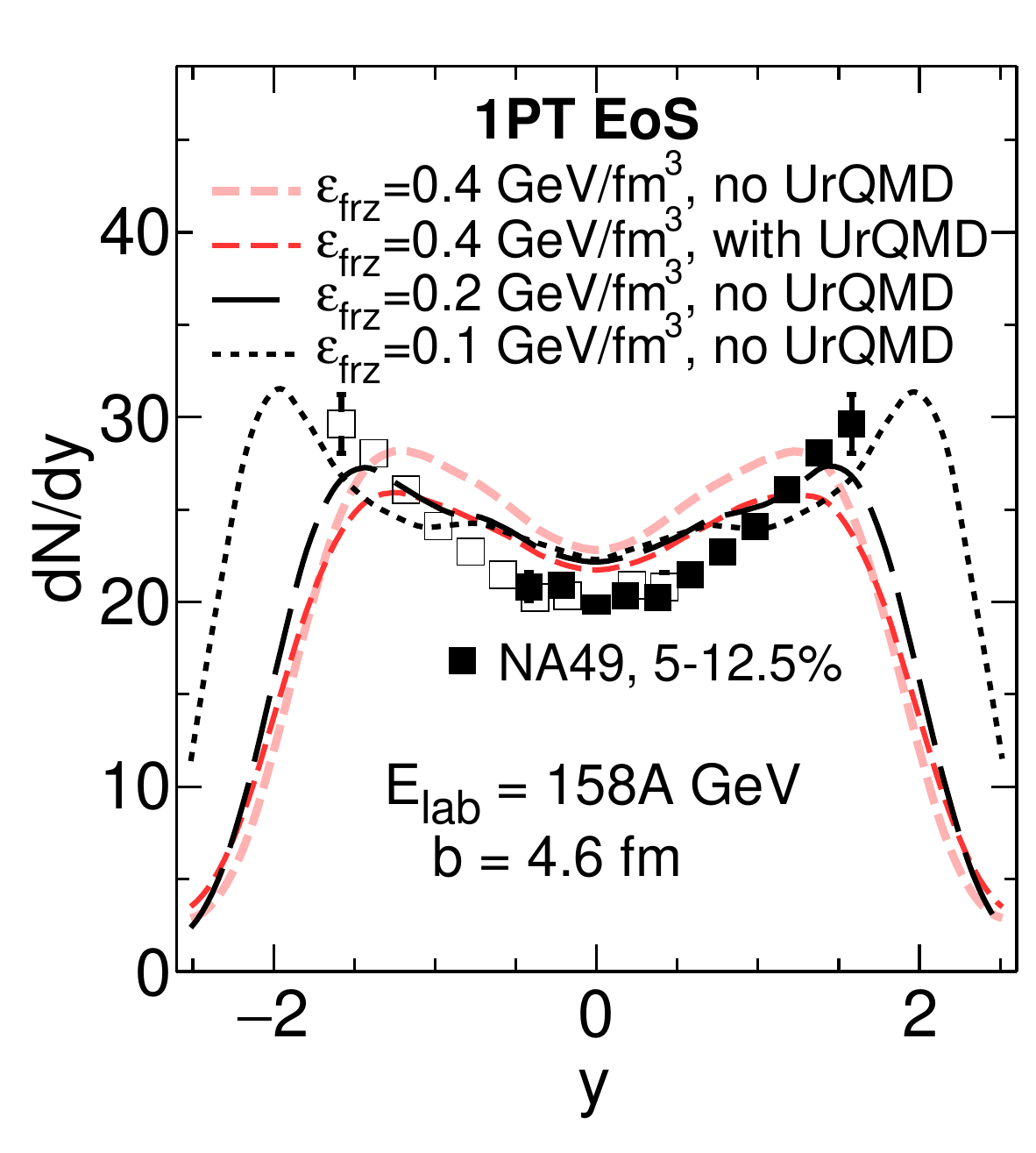}
  \caption{(Color online)
Rapidity distributions of net-protons in central ($b=$ 2.4 fm) 
Pb+Pb collisions at $E_{\rm lab}=$ 158$A$ GeV   
calculated with the crossover EoS (upper panel) and the 1PT EoS (lower panel). 
Results of the THESEUS simulations (without the subsequent UrQMD afterburner)
based on the 3FD calculations with different 
freeze-out energy densities $\varepsilon_{\rm frz}=$ 0.1, 0.2 and 0.4 GeV/fm$^3$ are shown. 
The conventional for the 3FD results with $\varepsilon_{\rm frz}=$ 0.4 GeV/fm$^3$ 
and the subsequent UrQMD afterburner are also presented.   
Experimental data are from the NA49 collaboration  \cite{Anticic:2010mp}.
}	
    \label{fig:dNdy_p_158AGeV}
\end{figure}

The effect of the late freeze-out on rapidity distribution of net-protons is demonstrated in 
Fig.  \ref{fig:dNdy_p_158AGeV}. The late freeze-out with $\varepsilon_{\rm frz}=$ 0.2 GeV/fm$^3$
reasonably well reproduces results for conventional freeze-out with 
the subsequent UrQMD afterburner in the midrapidity region. 
The reproduction for $E_{\rm lab}=$ 20$A$ GeV (not shown) is even better, as it can be expected 
from  Figs. \ref{fig:pT_p_20AGeV}. 
However, the freeze-out energy density $\varepsilon_{\rm frz}=$ 0.2 GeV/fm$^3$ is not that good in 
imitating the afterburner effect at forward/backward rapidities, see Fig.  \ref{fig:dNdy_p_158AGeV}. 
Below we use this late freeze-out with $\varepsilon_{\rm frz}=$ 0.2 GeV/fm$^3$ for calculations of light 
nuclei for all considered collision energies and centralities. 

As seen from Fig.  \ref{fig:dNdy_p_158AGeV}, 
the rapidity-integrated net-proton yield $\varepsilon_{\rm frz}=$ 0.1 GeV/fm$^3$ is visibly larger 
than that at higher $\varepsilon_{\rm frz}$ because the the rapidity distribution at 0.1 GeV/fm$^3$
extends to larger forward/backward repidities. 
The reason is that only participant test particles 
(at the Lagrangian step of the 3FD) are recoded in 
the 3FD output and thus transferred to the THESEUS. The spectator test particles%
\footnote{
i.e. those, the proper energy per baryon charge of which is less than the nucleon mass
}, 
containing bound nuclear matter, 
are omitted. At the late freeze-out, the participant region expands, involving more and more former
spectators. Therefore, the total number of net-proton participants increases.

\subsection{Rapidity distributions} 
  \label{Rapidity distributions}

\begin{figure*}[!tbh]
  \includegraphics[width=.93\textwidth]{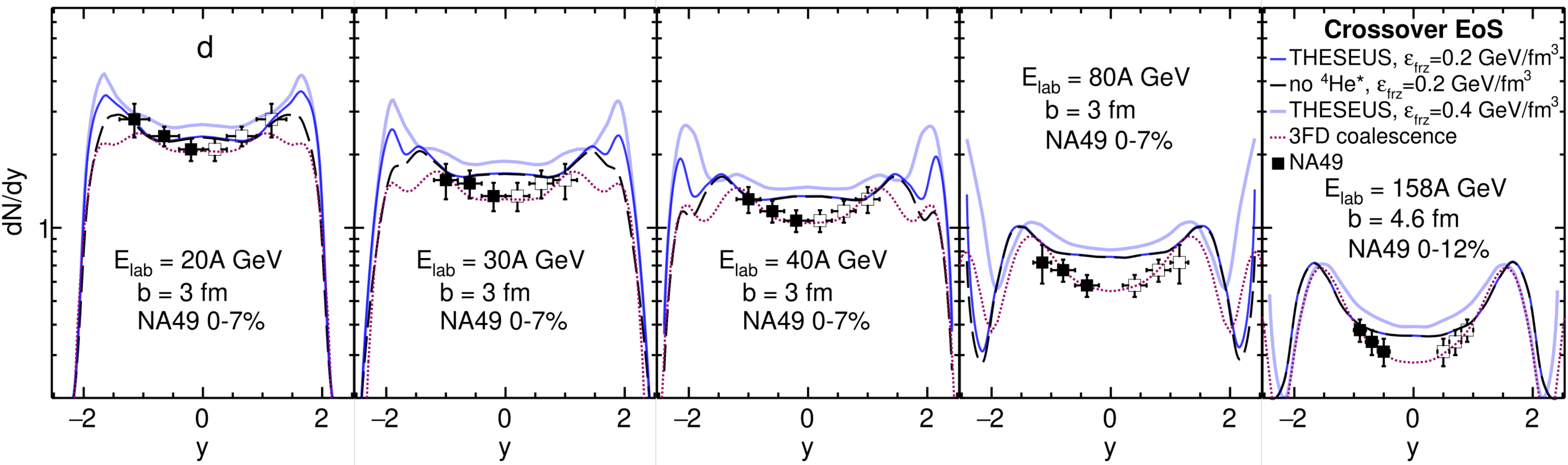}
  \includegraphics[width=.93\textwidth]{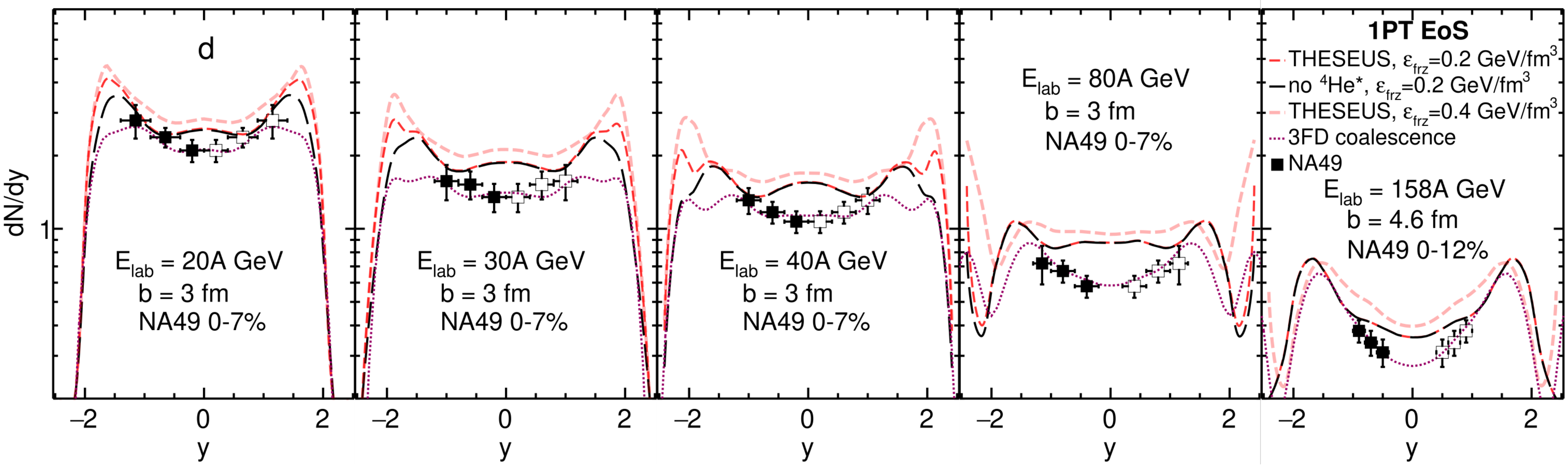}
  \caption{(Color online)
Rapidity distributions of deuterons in central 
Pb+Pb collisions at collision energies of $E_{\rm lab}=$ 20$A$-158$A$ GeV  
calculated with the crossover EoS (upper raw of panels) and the 1PT EoS (lower raw of panels). 
Results of THESEUS simulations with the conventional freeze-out, $\varepsilon_{\rm frz}=$ 0.4 GeV/fm$^3$, 
and the late freeze-out, $\varepsilon_{\rm frz}=$ 0.2 GeV/fm$^3$, are displayed. 
Results of the simulations without contribution of low-lying resonances of the $^4$He
and the 3FD coalescence results \cite{Ivanov:2017nae} are also presented. 
Experimental data are from the NA49 collaboration  \cite{Anticic:2016ckv}.
}	
    \label{fig:Dmix}
\end{figure*}
\begin{figure*}[!bth]
  \includegraphics[width=.93\textwidth]{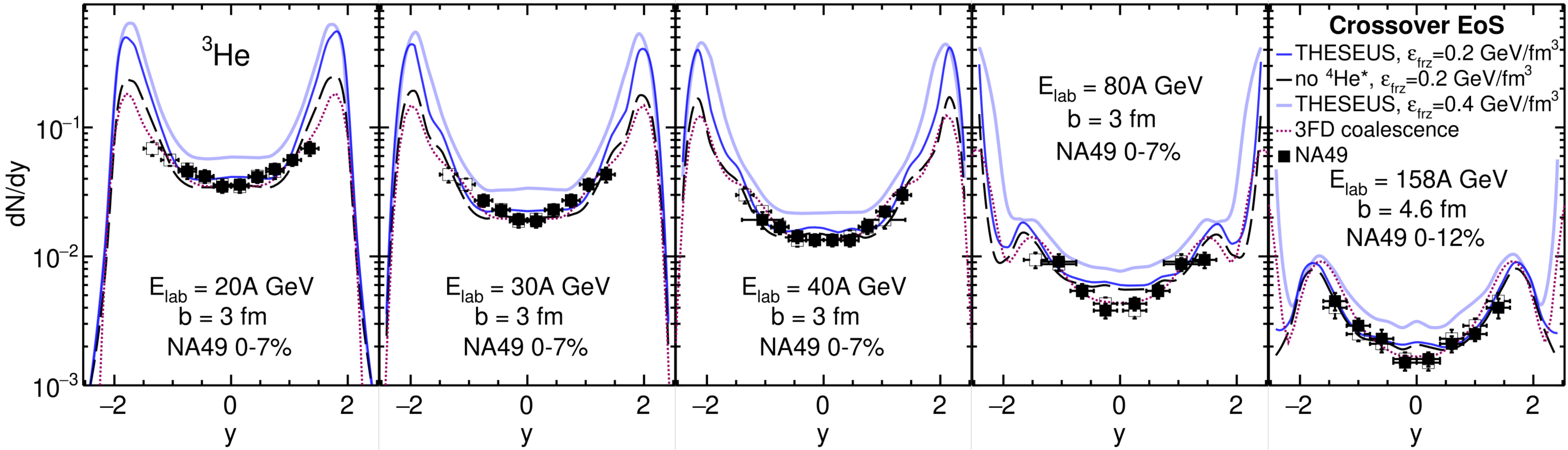}
  \includegraphics[width=.93\textwidth]{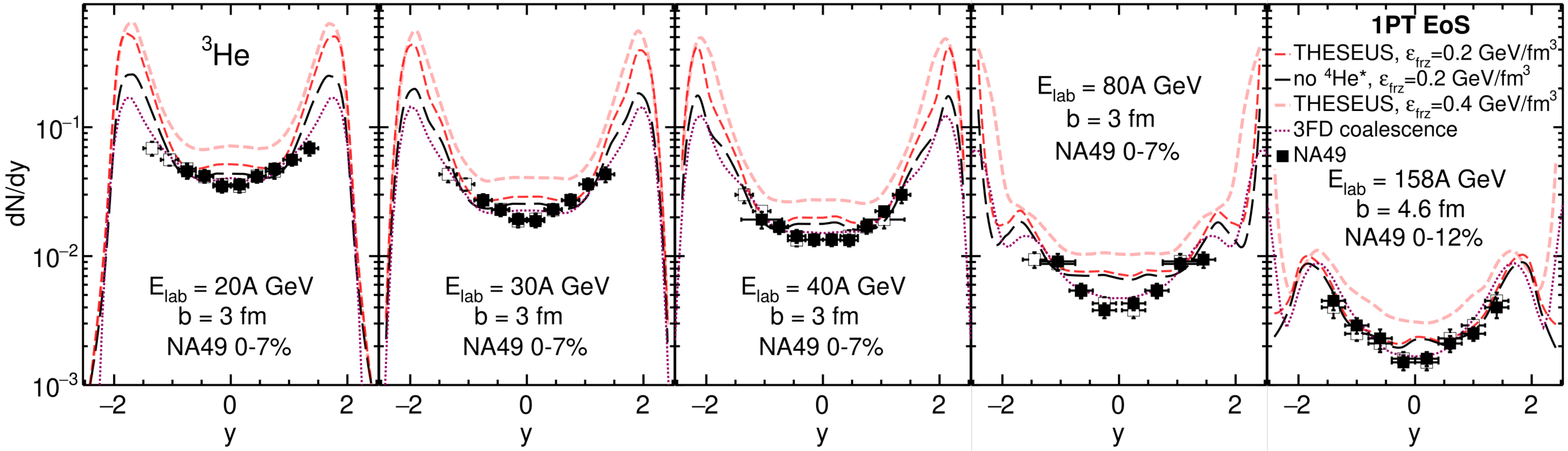}
  \caption{(Color online)
The same as in Fig. \ref{fig:Dmix} but for the $^3$He nuclei. 
}	
    \label{fig:He3mix}
\end{figure*}
\begin{figure*}[!tbh]
  \includegraphics[width=.93\textwidth]{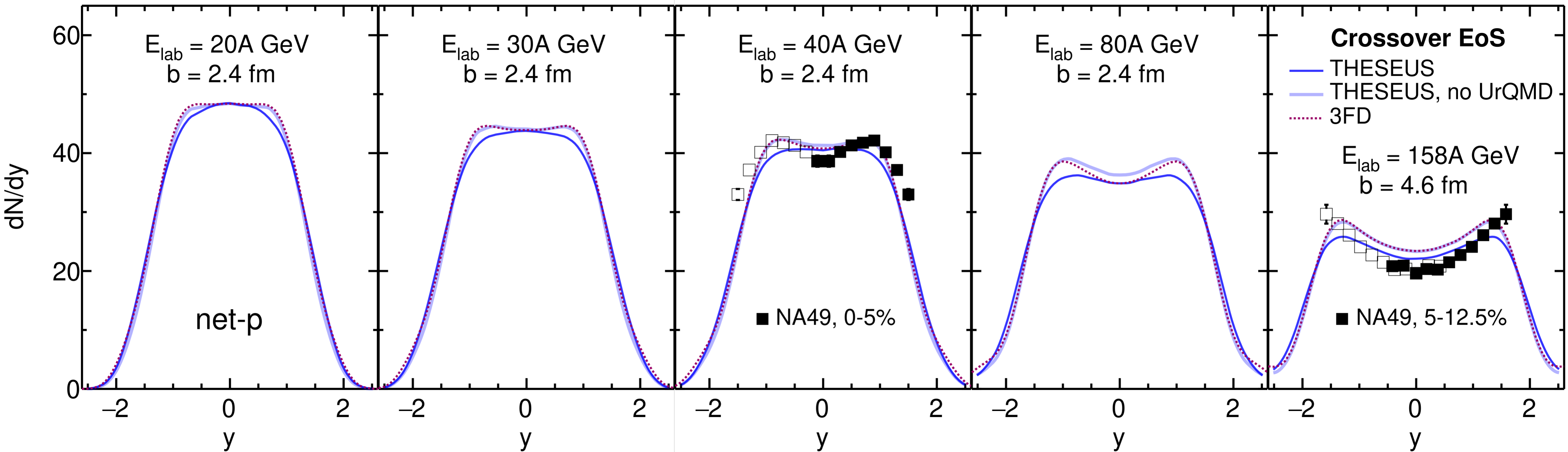}
  \includegraphics[width=.93\textwidth]{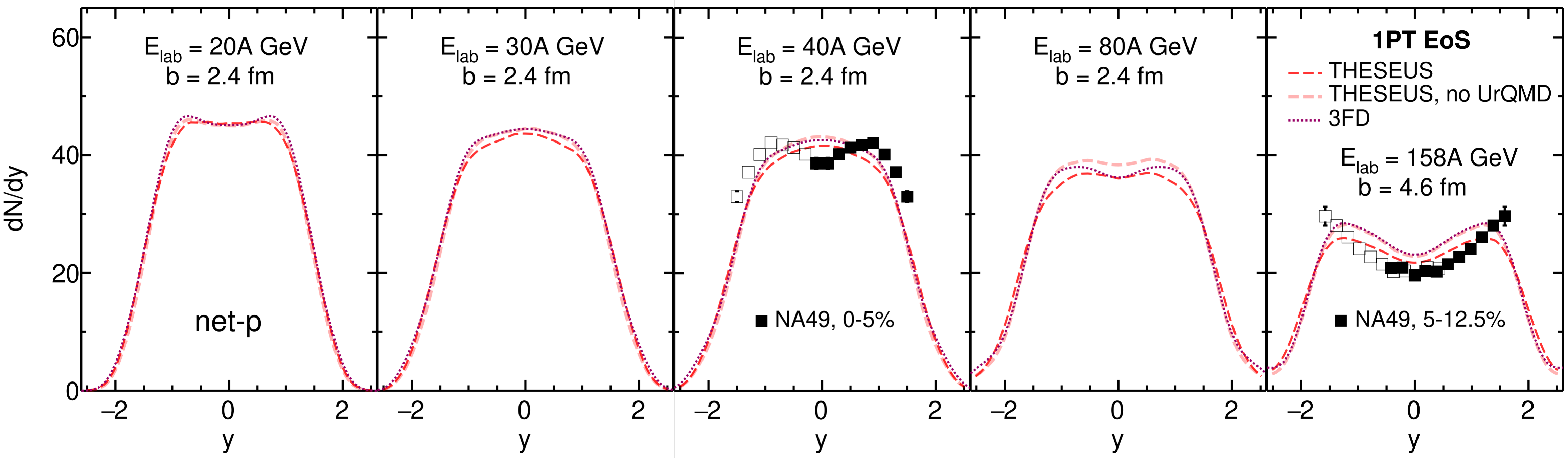}
  \caption{(Color online)
The same as in Fig. \ref{fig:Dmix} but for 
net-protons in central 
Pb+Pb collisions. 
Results of simulations without the UrQMD afterburner
and the 3FD results \cite{Ivanov:2017nae} are also displayed. 
Experimental data are from the NA49 collaboration  \cite{Anticic:2010mp}.
}	
    \label{fig:Pmix}
\end{figure*}

We start with analysis of rapidity distributions of light nuclei, see Figs. \ref{fig:Dmix} and \ref{fig:He3mix}.
We compare our results with NA49 data \cite{Anticic:2016ckv}, 
as well as with the 3FD coalescence results \cite{Ivanov:2017nae}. 
Let us remind that the light nuclei are simulated without the afterburner stage. 
To illustrate once again the expected effect of the afterburner stage, 
we present 
results of the simulations with conventional freeze-out, $\varepsilon_{\rm frz}=$ 0.4 GeV/fm$^3$, 
and late freeze-out, $\varepsilon_{\rm frz}=$ 0.2 GeV/fm$^3$, which imitate the afterburner stage.  
As seen, in the midrapidity region the THESEUS results 
systematically overestimate the data on light-nuclei yields. 
The late freeze-out somewhat improves agreement with the data but not completely. 
The extent of this agreement depends on the EoS. The crossover EoS results in better 
agreement with data than the 1PT EoS. 
It is surprising that reproduction of the $^3$He data turns out to be better than that 
of the data on deuterons, in spite of that $^3$He is a heavier nucleus.

The energy of $E_{\rm lab}=$ 80$A$ GeV drops out of the systematics. The disagreement with 
data at 80$A$ GeV is larger than at neighboring energies of 40$A$ and 158$A$ GeV,   
as if the clustering is additionally suppressed at 80$A$ GeV.  
The 3FD coalescence \cite{Ivanov:2017nae}, also presented in Figs. \ref{fig:Dmix} and \ref{fig:He3mix}
by short-dashed lines, much better 
reproduces the data because the coalescence coefficients were tuned for 
each collision energy and each light nucleus. 
Nevertheless, the THESEUS simulations result in good agreement 
with the dependence of light-nuclei production on the collision energy and light-nucleus mass. 
This agreement does not need any additional tunning parameters, i.e. in addition to those used for 
description of all other hadron yields, contrary to the 3FD coalescence.

Results of simulations without the contribution of low-lying resonances of the $^4$He
are also displayed Figs. \ref{fig:Dmix} and \ref{fig:He3mix} by long-dashed lines. 
The effect of low-lying resonances of the $^4$He system in the midrapidity region 
is small at the considered collision energies, 
as it has been already mentioned in ref. \cite{Kozhevnikova:2020bdb}. 
However, it is essential in the fragmentation regions. 
These feed-down contributions from decays of unstable $^4$He are compatible with 
the results obtained within the statistical model \cite{Vovchenko:2020dmv}.

For comparison, the rapidity distributions of net-protons are presented in Fig. \ref{fig:Pmix}. 
The net-protons are reproduced much better. The UrQMD afterburner slightly reduces net-proton
yield in the midrapidity region and drives it to even better agreement with data. This reduction 
rises with the collision energy increase. Still it 
is much smaller than that for light nuclei, i.e. the difference between calculations with 
$\varepsilon_{\rm frz}=$ 0.2 and 0.4 GeV/fm$^3$. The crossover EoS again results in better 
agreement with available data than the 1PT EoS. Apparently, small inconsistency with proton data
transforms in large inconsistency with data on light nuclei.

The rapidity distributions of net-protons and light nuclei are quite different. While the net-proton 
distributions at $E_{\rm lab}=$ 20$A$-80$A$ GeV reveal a peak or a shallow dip at midrapidities, 
the distributions of light nuclei demonstrate high maxima at forward/backward rapidities. These maxima
are formed because the matter at the perephery of colliding system 
(i.e. at forward/backward rapidities) is colder than in its center (i.e. in midrapidity) and 
hence the relative abundances of light nuclei become large. Only small humps in the midrapidity 
of light-nuclei distributions remind about midrapidity peaks in the net-protons.

\subsection{Transverse-momentum spectra} 
  \label{Transverse-momentum spectra}

\begin{figure*}[!tbh]
  \includegraphics[width=.93\textwidth]{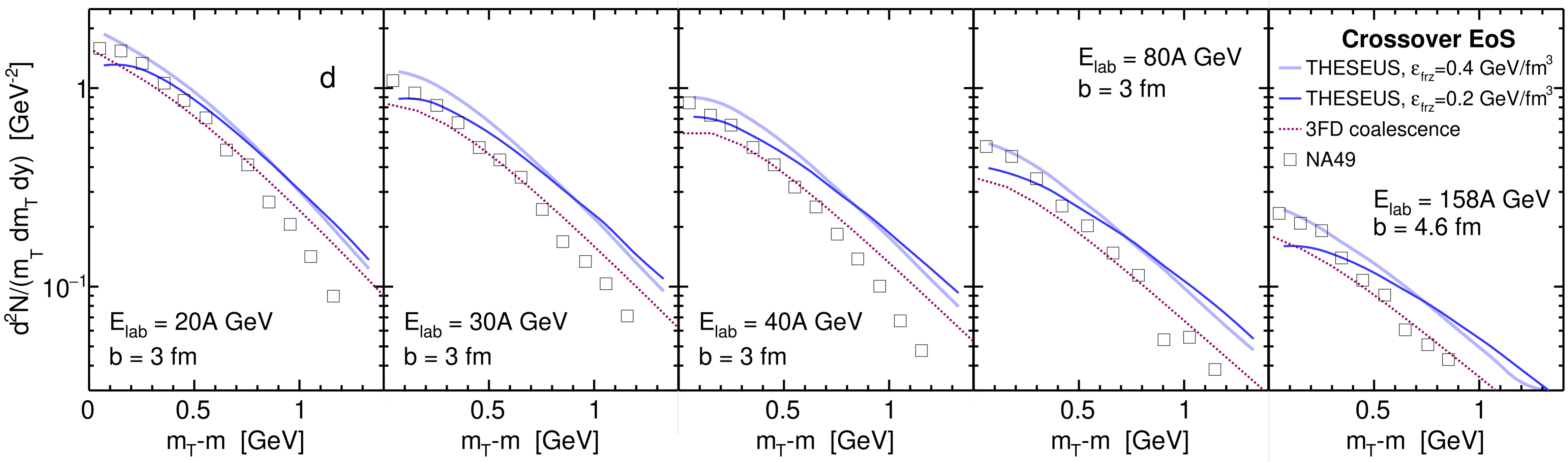}
  \includegraphics[width=.93\textwidth]{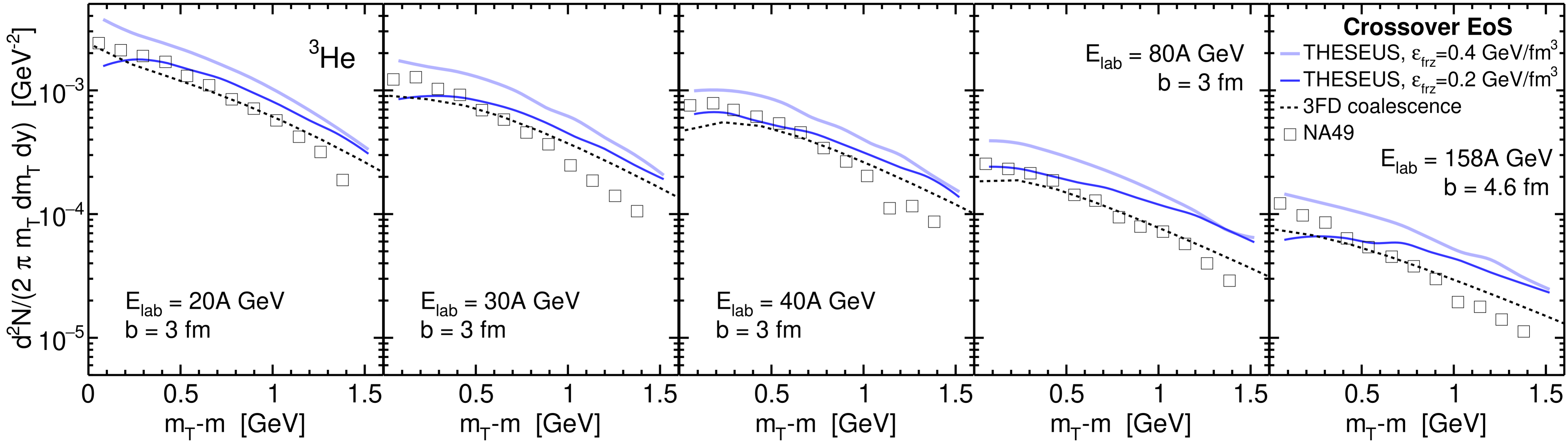}
  \caption{(Color online)
Transverse-mass spectra of deuterons (upper raw of panels) and tritons (lower raw of panels) in central 
Pb+Pb collisions at collision energies of $E_{\rm lab}=$ 20$A$-158$A$ GeV  
calculated with the crossover EoS. 
Results of THESEUS simulations with conventional freeze-out, $\varepsilon_{\rm frz}=$ 0.4 GeV/fm$^3$, 
and late freeze-out, $\varepsilon_{\rm frz}=$ 0.2 GeV/fm$^3$, are displayed. 
The 3FD coalescence results \cite{Ivanov:2017nae} are also shown.
NA49 data are from Ref. \cite{Anticic:2016ckv}.
}	
    \label{fig:mT_all_mix}
\end{figure*}
\begin{figure*}[bht]
  \includegraphics[width=.93\textwidth]{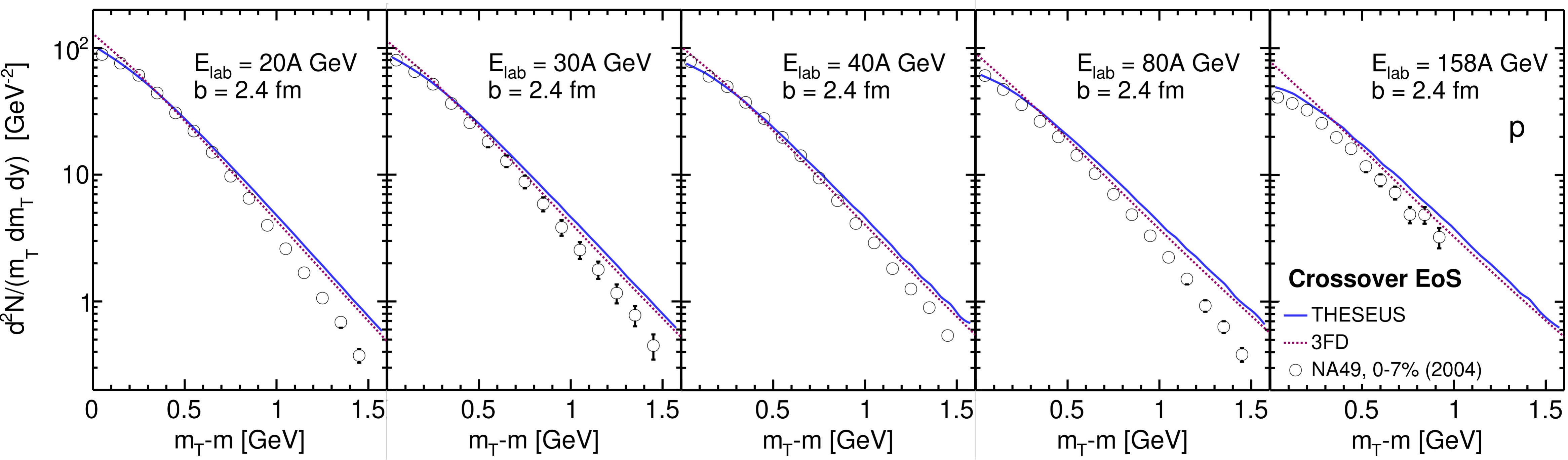}
  \includegraphics[width=.93\textwidth]{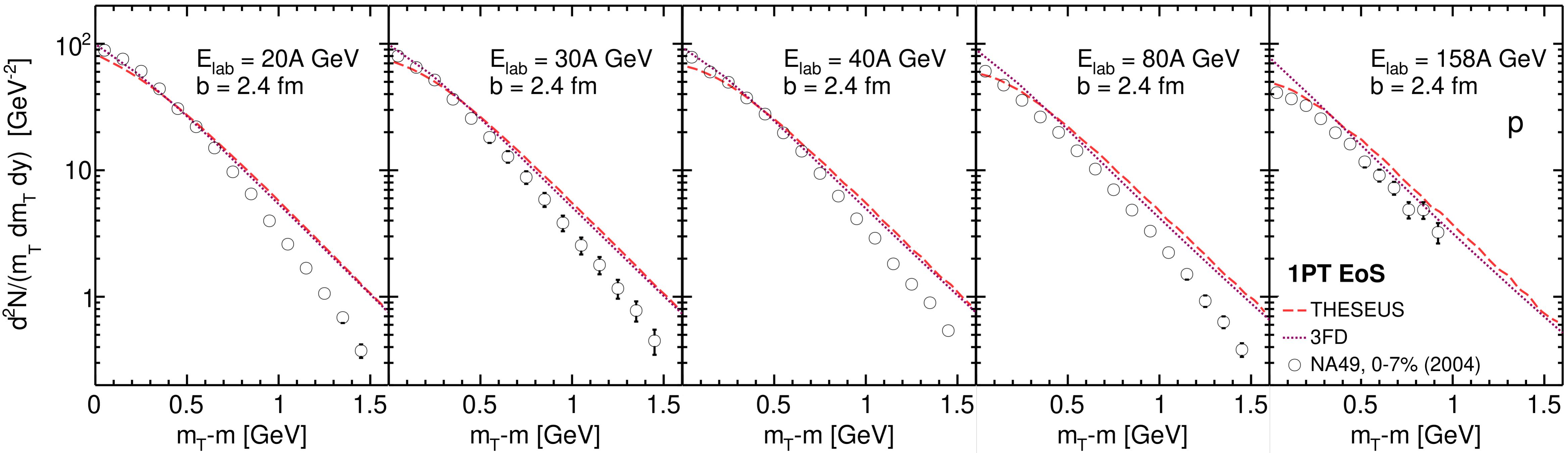}
  \caption{(Color online)
The same as in Fig. \ref{fig:mT_all_mix} but for protons 
calculated with the crossover EoS (upper raw of panels) and the 1PT EoS (lower raw of panels).  
Results of conventional THESEUS simulations (i.e. with the UrQMD afterburner)
and the 3FD results \cite{Ivanov:2017nae} are displayed. 
NA49 data are from Ref. \cite{NA49:2004iqm}.
}	
    \label{fig:mT_p_mix}
\end{figure*}

Transverse-mass spectra of deuterons and tritons at midrapidity in central 
Pb+Pb collisions at collision energies of $E_{\rm lab}=$ 20$A$-158$A$ GeV, 
measured by the NA49 collaboration \cite{Anticic:2016ckv},  
are compared with  the results of the THESEUS simulations in Fig. \ref{fig:mT_all_mix}.  
Only the calculations with the crossover EoS are shown because the 1PT scenario results in 
a similar picture, where agreement with data is even slightly worse. 
Again, results of THESEUS simulations with conventional freeze-out, $\varepsilon_{\rm frz}=$ 0.4 GeV/fm$^3$, 
and late freeze-out, $\varepsilon_{\rm frz}=$ 0.2 GeV/fm$^3$, are displayed. 
The 3FD coalescence results \cite{Ivanov:2017nae} are also shown.

The slopes of the 3FD-coalescence and THESEUS (with conventional freeze-out) spectra are 
very similar, in spite of being obtained within different approaches. 
This is because the coalescence and thermodynamical expressions for the light-nuclei yields 
are very similar except for the tunable coalescence coefficients
implied in the coalescence approach, which control the overall normalization. 
The agreement of these spectra with the NA49 data is far from being perfect. 
The normalization of the light-nuclei spectra is strongly overestimated
within the THESEUS with conventional freeze-out. Imitation of the afterburner
(THESEUS with $\varepsilon_{\rm frz}=$ 0.2 GeV/fm$^3$) somewhat improves the 
normalization at low $m_T-m$ but worsen agreement with the slopes. 
The overall normalization of the 3FD-coalescence spectra is better but this is achieved 
by tuning the coalescence parameters.

The late freeze-out makes slopes of the $m_T$ spectra less steep, which is an expected effect 
of the afterburner \cite{Song:2010aq}. This effect of the afterburner on proton spectra 
is demonstrated in Fig. \ref{fig:mT_p_mix}. For the protons, it is the UrQMD afterburner
after the conventional 0.4 GeV/fm$^3$ freeze-out rather than the late freeze-out for light nuclei. 
In the case of the late freeze-out,  
this flattening of the slope is a result of an increase of the radial-flow velocity over time.

The proton transverse-mass spectra at midrapidity in central 
Pb+Pb collisions at the same collision energies are presented in Fig. \ref{fig:mT_p_mix}.
Again, results of 3FD simulations are also displayed. 
As seen, the afterburner (THESEUS curves in Fig. \ref{fig:mT_p_mix}) 
improves agreement with the NA49 data \cite{NA49:2004iqm} at low $m_T-m$ as compared with the 3FD, 
but the slopes disagree with the data. Though, this disagreement is much smaller than that 
for light nuclei in Fig. \ref{fig:mT_all_mix}.

The calculated $^3$He spectra are closer to the data than the deuteron data, which is again surprising. 
The spectra slopes are better reproduced at lower energies. Together with better agreement with 
rapidity distributions of light nuclei at lower energies, 
this may suggest that the THESEUS is more suitable for 
simulating light nuclei at NICA and FAIR energies.

Again we may conclude that small disagreement with proton data
transforms into a large disagreement with data on light nuclei. 
In particular, slightly better reproduction of the proton data within the crossover scenario, 
as compared with the 1PT one, see Figs. \ref{fig:Pmix} and \ref{fig:mT_p_mix}, 
results in noticeably better agreement with data on light nuclei, cf. Fig. \ref{fig:He3mix}.

The 3FD predictions overestimate the high-$m_T$ ends of the spectra
because of finiteness of the considered system. 
Even abundant hadronic probes become rare at
high momenta. Therefore, their treatment on the basis
of grand canonical ensemble results in overestimation of
their yields. Moreover, the more rare probe is, the stronger its high-$p_T$ end of the spectrum is
suppressed due to restrictions of the canonical ensemble. 
Therefore, the light-nuclei high-$p_T$ spectra are stronger overestimated than the proton spectra. 
The UrQMD afterburner, as it is implemented in the THESEUS \cite{Kozhevnikova:2020bdb,Batyuk:2016qmb}, 
does not improve the high-$p_T$ description. 
The reason is that the grand canonical distributions are sampled in the particlization procedure, 
rather than the canonical or micro-canonical ones. Thus, the high-$p_T$ overestimation persists. 
Of course, it is difficult to indicate how much of this overestimation is due to the grand canonical treatment, and not to the shortcomings of the model.

We did not tune the 3FD model to reproduce the data on light nuclei, in particular, the 
$m_T$ spectra. The poor agreement with the data on the $m_T$ spectra is the price paid 
for the intention to reproduce numerous data in a wide range of collision 
energies with the same set of parameters described in Ref. \cite{Ivanov:2013wha}.

\subsection{Yield ratios of light-nuclei } 
  \label{ratios}

\begin{figure}[!tbh]
  \includegraphics[width=.43\textwidth]{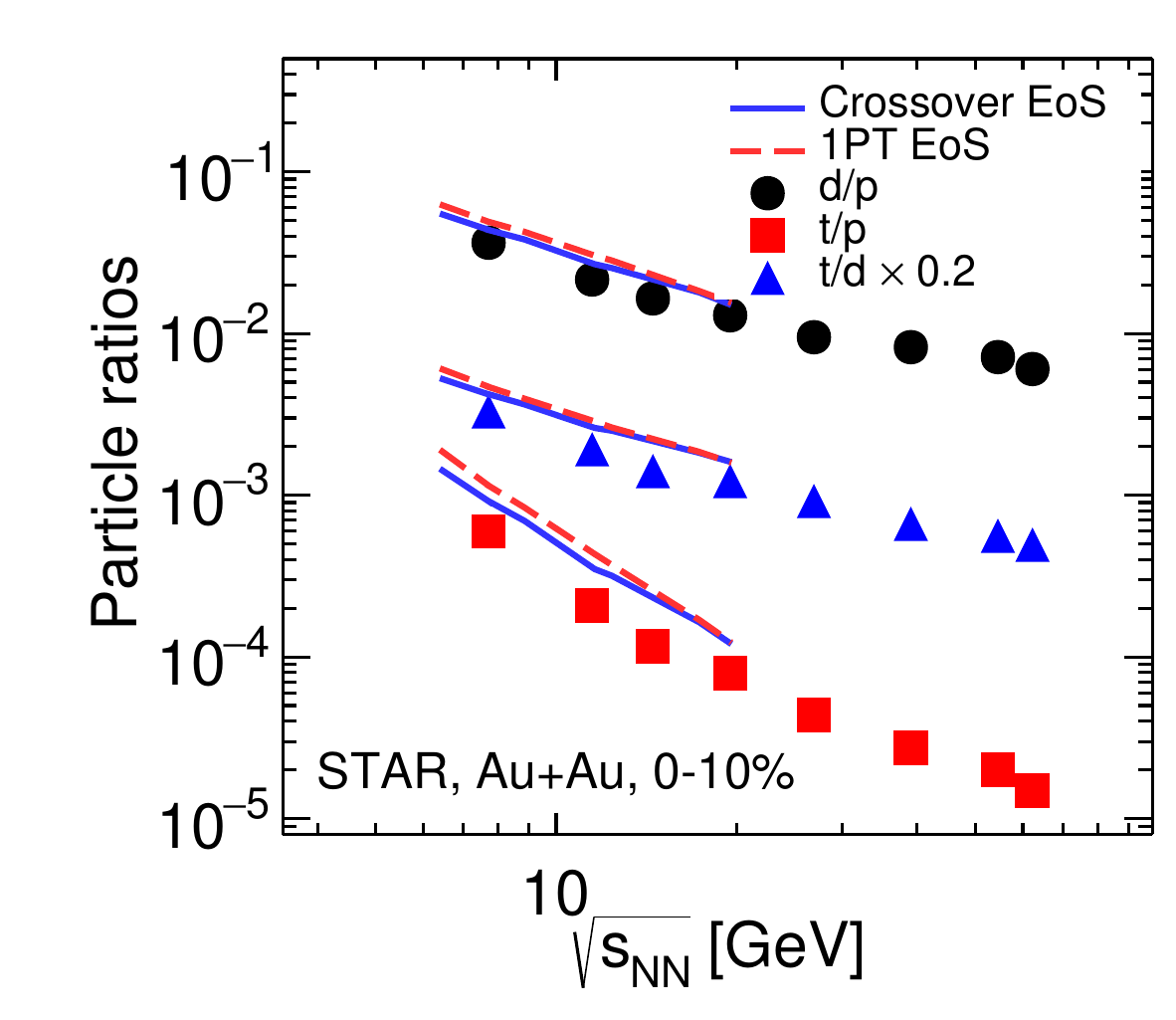}
  \caption{(Color online)
Energy dependence of $d/p$, $t/p$,
and $t/d$  midrapidity ratios for central (0-10\%) Au+Au collisions.
Simulations were performed at $b=$ 4 fm  for Au+Au and at $b=$ 3 fm 
for Pb+Pb in rapidity bin $|y|<0.5$.  
Results with the crossover and 1PT EoS's are presented. 
Results of the calculations are compared with STAR data \cite{STAR:2022hbp} for 
central (0--10\%) Au+Au collisions.
}	
    \label{fig:ParticleRatios}
\end{figure}

Energy dependence of $d/p$, $t/p$, and $t/d$ midrapidity ratios for central collisions 
are presented in Fig. \ref{fig:ParticleRatios}. Protons in these ratios do not contain
feed-down from weak decays, in accordance with the experimental procedure \cite{STAR:2019sjh,STAR:2022hbp}.  
As can be seen, the model reproduces the energy dependence of experimental data \cite{STAR:2022hbp} 
but systematically overestimates the values of these ratios. 
This reproduction is similar to that within the statistical model \cite{STAR:2022hbp}.

\begin{figure}[!tbh]
  \includegraphics[width=.43\textwidth]{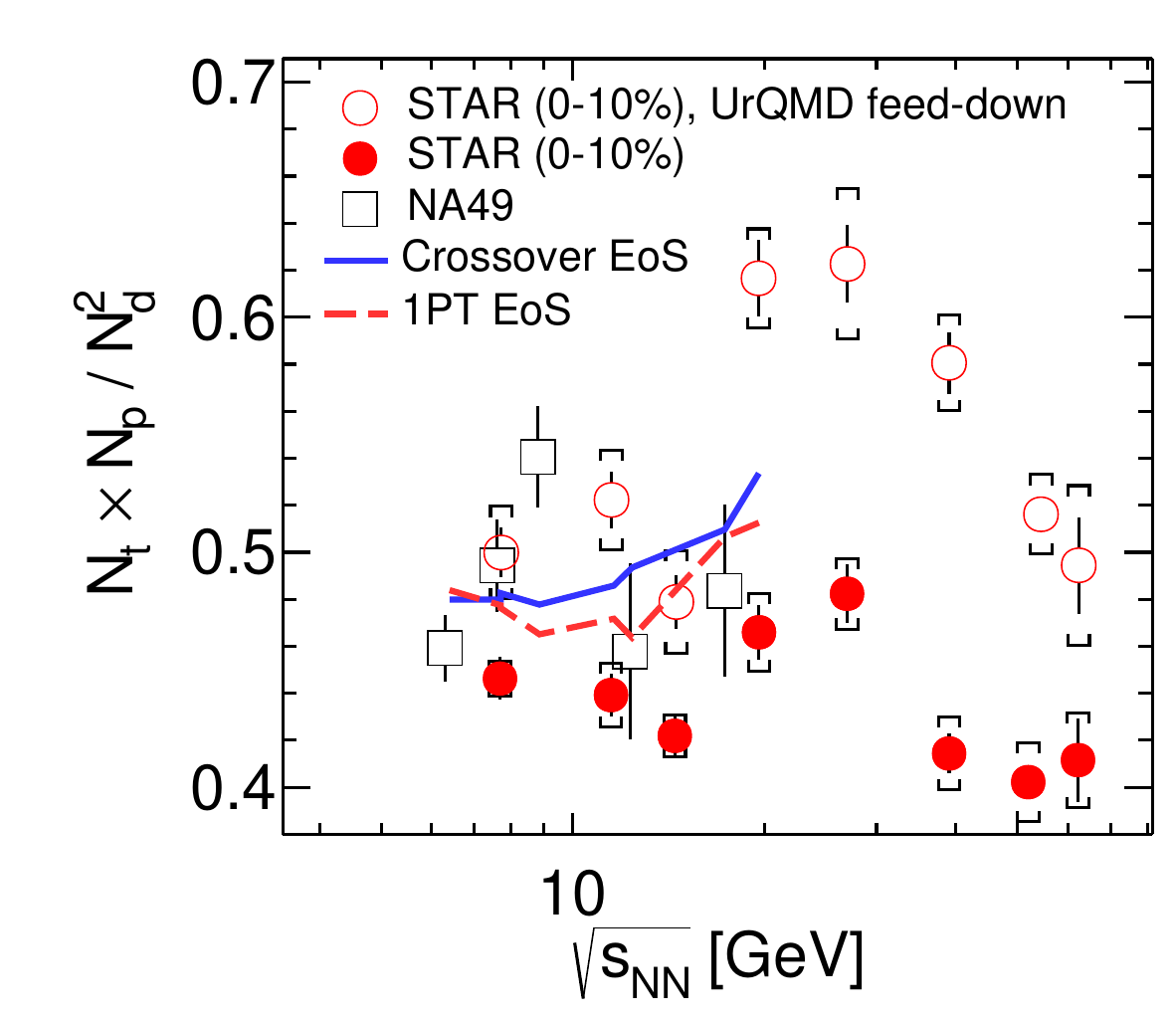}
  \caption{(Color online)
Energy dependence of the midrapidity light-nuclei-yield ratio $N(t)\times N(p)/N^2(d)$ in central Au+Au 
and Pb+Pb 
collisions. 
Simulations were performed at $b=$ 4 fm  for Au+Au,
at $b=$ 3 fm ($\sqrt{s_{NN}}<$ 17.4 GeV) and $b=$ 4.6 fm ($\sqrt{s_{NN}}=$ 17.4 GeV) for Pb+Pb
in rapidity bin $|y|<0.5$.  
$N(p)$ is related to 
protons without feed-down from weak decays. 
Results of the calculations are compared with STAR preliminary data \cite{Zhang:2020ewj} (open circles), 
where weak-decays feed-down into proton yield was determined by means of the UrQMD simulations, 
and final data \cite{STAR:2022hbp} (filled circles), 
with the weak-decays feed-down determined by experimental means, for 
central (0--10\%) Au+Au collisions.  
The experimental results extracted from the NA49 data on Pb+Pb collisions
(0--7\% at 20$A$-80$A$ GeV and 0--12\% at 158 $A$ GeV)
\cite{Anticic:2016ckv} are also displayed (open boxes).
}	
    \label{fig:NtNpdivN2d}
\end{figure}
\begin{figure*}[!tbh]
  \includegraphics[width=.64\textwidth]{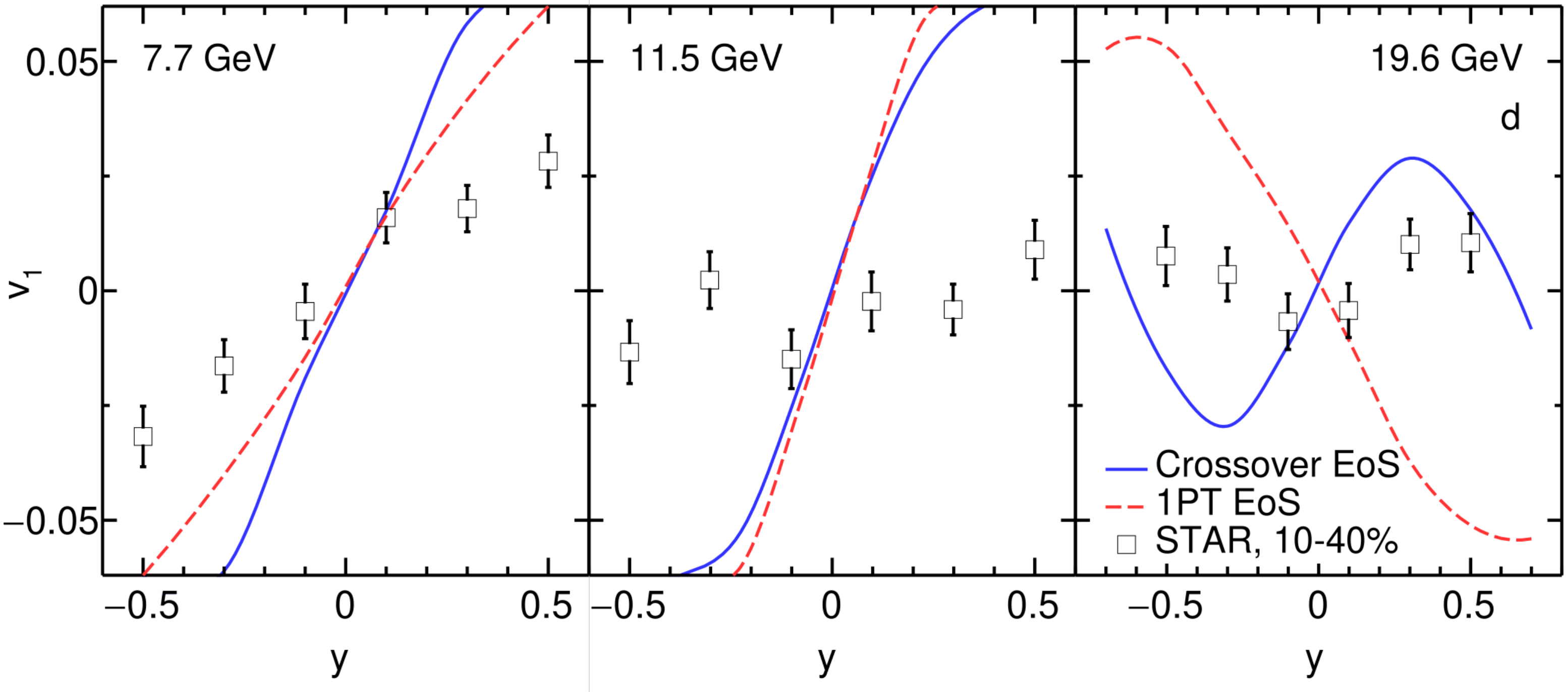}
  \includegraphics[width=.64\textwidth]{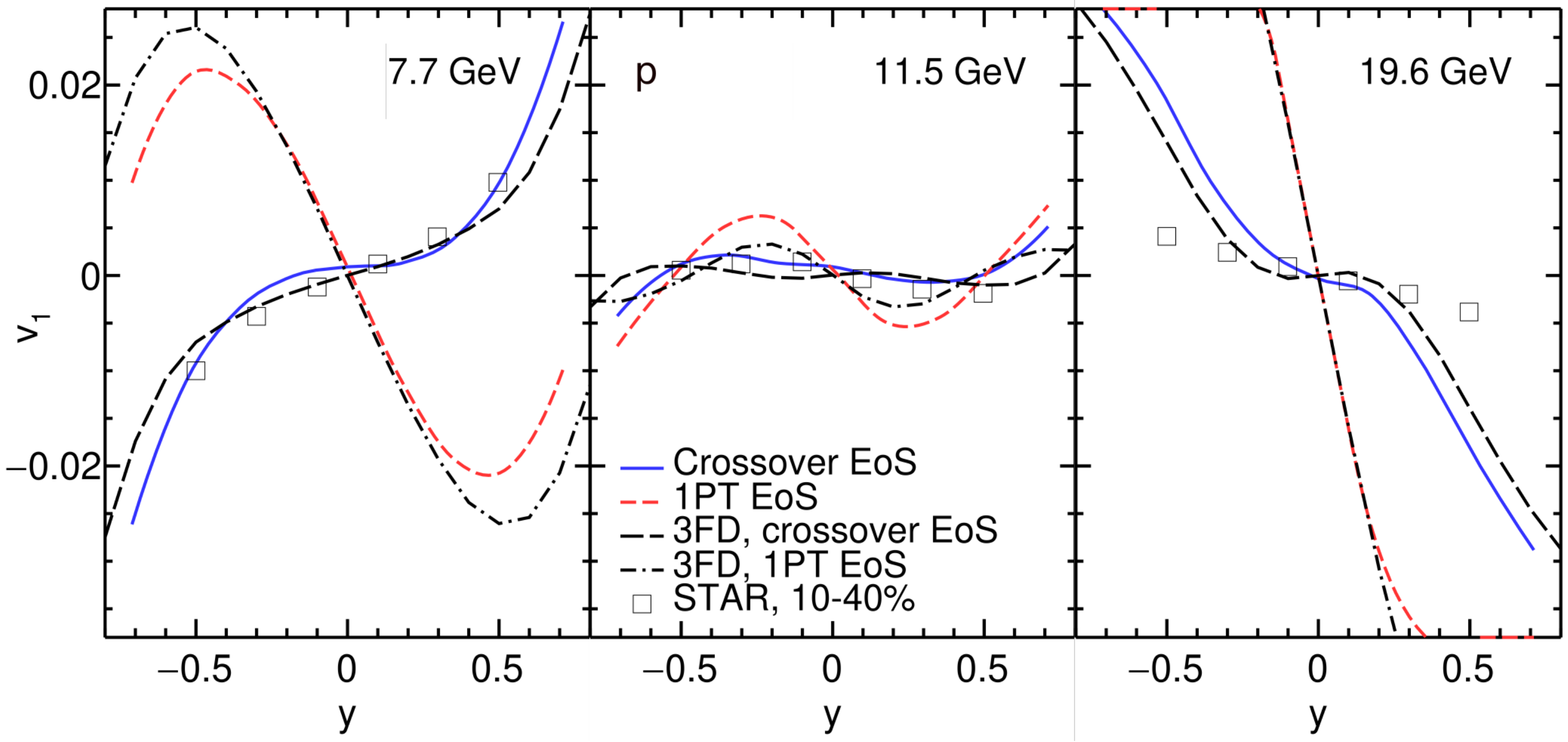}
  \caption{(Color online)
Directed flow of  deuterons (upper raw of panels) and protons (lower raw of panels)  
as function of rapidity in semicentral ($b=$ 6 fm) 
Au+Au collisions at  collision energies of $\sqrt{s_{NN}}=$ 7.7, 11.5 and 19.6 GeV 
calculated with the crossover and 1PT EoS's.  
Results of the 3FD simulations \cite{Ivanov:2014ioa}, i.e. without the UrQMD afterburner, are also 
presented for protons. 
Experimental STAR data for deuterons are from Ref.  \cite{STAR:2020hya} and 
for protons are from Ref.  \cite{STAR:2014clz}.
}	
    \label{fig:v1_d_frz02}
\end{figure*}

The yield ratio of light nuclei, $N_t N_p/N^2_d$, has been suggested as a probe to the neutron density fluctuations associated with the first-order phase
transition \cite{Sun:2017xrx,Sun:2018jhg}.
Later it was also associated with the possible critical point of the hot and
baryon-rich QCD matter \cite{Shuryak:2019ikv,Shuryak:2020yrs,Sun:2020zxy}. Near the critical point, 
this ratio increases monotonically with the nucleon density correlation length \cite{Sun:2020zxy}, 
besides, production of $^3$He 
may increase because of enhanced preclustering and subsequent decay of $^4$He-like clusters 
\cite{Shuryak:2019ikv,Shuryak:2020yrs}.
In its turn, this may result in a maximum in the $N_t N_p/N^2_d$ ratio near the critical point. 
Recent data on this ratio \cite{STAR:2022hbp} show a non-monotonic behavior with a
peak located around 20--30 GeV (see Fig. \ref{fig:NtNpdivN2d}), which might 
indicate passing through 
either the first-order phase transition or critical point at this collision energy.

Energy dependence of the midrapidity $N(t)\times N(p)/N^2(d)$ ratio  in central Au+Au 
and Pb+Pb collisions is presented in Fig. \ref{fig:NtNpdivN2d}.  
Simulations were performed at $b=$ 4 fm  for Au+Au,
at $b=$ 3 fm ($\sqrt{s_{NN}}<$ 17.4 GeV) and $b=$ 4.6 fm ($\sqrt{s_{NN}}=$ 17.4 GeV) for Pb+Pb
in rapidity bin $|y|<0.5$ with the crossover and 1PT EoS's. 
The proton yields does not include contribution from the weak-decay feed-down. 
This weak-decay feed-down was determined by the UrQMD simulation at the afterburner stage, 
i.e. in the same way as the proton feed-down correction was done in the preliminary 
STAR data \cite{Zhang:2020ewj}. 
In the final STAR data \cite{STAR:2022hbp} (filled circles in Fig. \ref{fig:NtNpdivN2d}), 
the feed-down correction was done by experimental means. 
The STAR Collaboration concludes that the UrQMD
simulation underestimates the proton feed-down contributions from weak decays.
If so, our simulations suffer from the same shortcoming of the UrQMD. 
Therefore, we also present these preliminary data \cite{Zhang:2020ewj} 
in Fig. \ref{fig:NtNpdivN2d} for comparison.

The calculated $N(t)\times N(p)/N^2(d)$ ratios overestimate the final STAR data both for
the crossover and 1PT EoS's. At the same time they are, as a rule, below preliminary STAR data. 
Therefore, the overestimation can be related to the aforementioned shortcoming of the UrQMD. 
The calculated ratios show an increase as the energy approaches 20 GeV, in spite of absence of 
the critical point in the considered EoS's. This is not an effect of a special tune of the model parameters. 
As described above, there were no such special tune. 
This increase can be also an artifact of the 
underestimation of the proton feed-down contributions from weak decays at the UrQMD stage. 
Indeed, from Fig. \ref{fig:NtNpdivN2d} we can see the difference between the 
preliminary and final STAR data becomes larger at $\sqrt{s_{NN}}\geq$ 20 GeV as compared with that 
at lower energies.

We may conclude that 
the feed-down contributions of hyperon weak decays  
should be carefully subtracted from the proton yield 
in order the calculated $N(t)\times N(p)/N^2(d)$ ratio can 
serve as probe of production characteristics of light
nuclei and the structure of the QCD phase diagram.
The UrQMD is not quite accurate for such subtraction. 
A similar conclusion was made in Ref. \cite{Zhao:2022xkz}.

\subsection{Directed flow} 
  \label{Directed flow}

The directed flow ($v_1$) is one of the most delicate characteristics of the heavy-ion collisions. 
Nevertheless, we  calculated the deuteron $v_1$ relying on relatively successful description of 
the proton $v_1$ within the 3FD model \cite{Ivanov:2014ioa}. 
This is a straightforward calculation because light nuclei are treated on the equal basis with other
hardrons in the present approach. 

The directed flow of deuterons (upper raw of panels in  Fig. \ref{fig:v1_d_frz02}), 
calculated for late freeze-out (i.e. $\varepsilon_{\rm frz}=$ 0.2 GeV/fm$^3$)
to imitate the afterburner effect,   
is shown and compared with  STAR  data  \cite{STAR:2020hya} in Fig. \ref{fig:v1_d_frz02}. 
The directed flow of protons (lower raw of panels in  Fig. \ref{fig:v1_d_frz02}) is also presented 
for comparison. 
In these simulations, experimental acceptance \cite{STAR:2020hya}
was used: 0.4 $< p_T <$ 2.0 GeV/c for protons and 0.8 $< p_T <$ 4.0 GeV/c for deuterons. 
To illustrate the effect of the afterburner, results of the 3FD simulations of protons $v_1$
are displayed. As seen, the  afterburner insignificantly changes proton $v_1$. 
The  ``afterburner'' effect in deuteron $v_1$ (not displayed) is sightly stronger but 
still not dramatic.

The directed flow of  deuterons is quite different from the proton one. 
As seen, the deuteron $v_1$ is stronger than the proton $v_1$. 
Even signs of the midrapidity slopes of deuteron and proton $v_1(y)$ are not always the same. 
The crossover and 1PT EoS's predict different $v_1$, which is not surprising because the 
proton $v_1(y)$ are also very different for these EoS's.
While the crossover EoS well reproduces the data on the proton $v_1$, 
overall reproduction of the deuteron data is much worse than that for protons. 
Nevertheless, the order of magnitude of the deuteron $v_1$ is comparable with the data, except for 
the energy of 11.5 GeV, where the deuteron $v_1$ collapses similarly to the proton $v_1$. 

Contrary to our naïve expectation,  
the reproduction of the proton directed flow does not guarantee a good description 
of that for light nuclei. The nucleon directed flow, represented by the proton one, measures 
azimuthal asymmetry of the baryon current  
because all baryonic resonances decay into nucleons after the freeze-out. 
The light-nuclei directed flow does not contain the contribution of the baryonic resonances 
in accordance to the thermodynamic approach.  
It additionally depends on the azimuthal asymmetry of the temperature and baryon density 
because denser and colder regions give larger contribution to the light-nuclei production. 
All these make the light-nuclei directed flow and even its midrapidity slope 
different from those of proton. 
In subsect. \ref{Rapidity distributions}, it has already been discussed how inhomogeneity 
of the temperature distribution along the beam direction 
makes rapidity distributions of net-protons and light nuclei quite different.

The afterburner stage may essentially change the light-nuclei flow
because of decays and re-production of light nuclei, as it is, e.g., realized in the 
SMASH \cite{Oliinychenko:2018ugs,Staudenmaier:2021lrg} and PHQMD \cite{Glassel:2021rod,Bratkovskaya:2022vqi} 
transport models. These processes may result in deviation from the kinetic equilibrium. 
Our imitation of the afterburner by means of the late freeze-out does not violate 
the kinetic equilibrium.

\section{Summary}
\label{Summary}

Simulations of the light-nuclei production 
in relativistic heavy-ion collisions within the updated THESEUS \cite{Kozhevnikova:2020bdb}
event generator were performed for Pb+Pb and Au+Au collisions in the collision energy range 
of $\sqrt{s_{NN}}=$  6.4--19.6 GeV. 
The results were compared with available data from the 
NA49 and STAR collaborations.
The updated THESEUS treats the light-nuclei production 
within the thermodynamical approach on the equal basis with hadrons.
The only additional parameter related to the light nuclei is the energy density of 
the late freeze-out, $\varepsilon_{\rm late \: frz}=$ 0.2 GeV/fm$^3$, which is the same 
for all collision energies, centralities and combinations of colliding nuclei. 
For comparison, the conventional freeze-out energy density in 3FD is 
$\varepsilon_{\rm frz}=$ 0.4 GeV/fm$^3$ that is also universal. 
This late freeze-out imitates the afterburner stage of the collision because the 
light nuclei are not subjected to the UrQMD afterburner. 
$\varepsilon_{\rm late \: frz}$ is not a free parameter, it is chosen from the condition  
of the best reproduction of the proton $p_T$ spectrum after the UrQMD afterburner 
by the spectrum at the late freeze-out without the afterburner.

The updated generator revealed not perfect, but a reasonable reproduction of 
the data on bulk observables of the light nuclei, 
especially the functional dependence on the collision energy  
and light-nucleus mass. 
It is important that this reproduction is achieved 
with a single universal additional parameter related to late freeze-out. 
The collective directed flow of light nuclei turns out to be more involved. 
Apparently, it requires an explicit treatment of the afterburner evolution of 
light nuclei with due account of violation of the kinetic equilibrium.

Various ratios, $d/p$, $t/p$, $t/d$, and $N(t)\times N(p)/N^2(d)$  
were also considered. We conclude that 
the feed-down contributions of weak decays  
should be carefully subtracted from the proton yield 
in order the calculated $N(t)\times N(p)/N^2(d)$ ratio can 
serve as probe of production characteristics of light
nuclei and the structure of the QCD phase diagram.
The UrQMD is not quite accurate for such subtraction.

Imperfect reproduction of the light-nuclei data leaves room for medium effects in production of light nuclei, 
which were advanced in Refs. \cite{Bastian:2016xna,Ropke:2017dur}, see also  \cite{Donigus:2022xrq}.

\begin{acknowledgments}

We are grateful to David Blaschke for convincing us to apply the thermodynamic approach 
to modeling the light-nuclei production in heavy-ion collisions. 
We are especially grateful to Iurii Karpenko, without his help and expertise 
this work would hardly have been possible. 
Useful discussions with V. Kireyeu are gratefully acknowledged. 
This work was carried out using computing resources of the supercomputer "Govorun" at JINR. 

\end{acknowledgments}


\end{document}